\documentclass[9pt,twocolumn,twoside]{osajnl}

\journal{ol} 

\setboolean{shortarticle}{true} 

\ifthenelse{\boolean{shortarticle}}{\colorlet{color2}{color2b}}{\colorlet{color2}{color2}} 

\title{Spin Hall effect and circular birefringence in polymers}

\author[1]{Osamu Takayama}
\author[2,3]{Graciana Puentes}

\affil[1]{Department of Photonics Engineering, Technical University of Denmark, Ørsteds Plads, Building 343,
DK-2800 Kgs. Lyngby, Denmark}
\affil[2]{Universidad de Buenos Aires. Facultad de Ciencias Exactas y
Naturales. Departamento de F\'{\i}sica. Buenos Aires, Argentina.}
\affil[3]{CONICET-Universidad de Buenos Aires. Instituto de F\'{\i}sica de Buenos Aires (IFIBA). Buenos Aires, Argentina.}

\affil[*]{Corresponding authors: gracianapuentes@gmail.com, otak@fotonik.dtu.dk}

\dates{Compiled \today}

\ociscodes{(260.0260) Physical optics; (260.5430) Polarization; (260.1440) Birefringence.}

\doi{\url{http://dx.doi.org/10.1364/ao.XX.XXXXXX}}

\begin{abstract}
We demonstrate  experimentally the fine lateral circular birefringence of a tunable birefringent polymer, the first example of the spin-Hall effect of light in a polymeric material. We report experimental observations of this effect using polarimetric techniques and quantum-weak-measurement techniques, reporting a weak amplification factor of 200. 
\end{abstract}

\setboolean{displaycopyright}{true}

\begin{document}

\maketitle

Spin-Orbit Interactions (SOI) of light refer to the coupling of different internal degrees of freedom of the radiation field, such as polarization and spatial degrees of freedom, as a result of propagation of light in different media. SOIs of light have recently attracted attention in a number of fast growing fields, ranging from photonics, plasmonics, nano-optics and quantum optics to meta-materials \cite{1}. Most SOI effects originate from space- or
wavevector-variant geometric phases and result in spin-dependent redistribution of light intensity \cite{1}. When the system has
cylindrical symmetry with respect to the z-axis, SOIs produce spin-to-orbital angular momentum conversion, i.e., generation
of a spin-dependent vortex in the z-propagating light \cite{1,2,3,4,5,6,7,8,9,10,11,12,13,14}. If the cylindrical symmetry is broken, SOIs bring about the spin-Hall effect of light, i.e., a spin-dependent transverse y-shift of light intensity \cite{12,13,14,15,16,17,22,23,24}. An
example of the latter effect is the so-called transverse Imbert–Fedorov (IF) beam shift or spin Hall effect of light, which occurs when a paraxial optical
beam is reflected or refracted at a plane interface \cite{22,23,24}. 

Up to present, spin-Hall effect has been observed for various material and geometrical settings, such as gratings \cite{3,9}, liquid crystals \cite{4}, dielectric spheres \cite{10}, metal \cite{PRA2012} and magnetic \cite{APL2012} films, uniaxial birefringent crystals \cite{Optica}, graphene \cite{APL2013}.  Furthermore, recently plasmonic \cite{Science2013} and dielectric \cite{LSA2015} meta-surfaces \cite{Nanophotonics2017}, as well as by hyperbolic metamaterials \cite{NatureCommun2014}, have been intensively studied due to their ability to manipulate polarization states by the design of artificial surface structures in subwavelength scale. Photonic spin-Hall effect offers device applications ranging from spin-dependent beam splitters \cite{Nanophotonics2017} to surface sensors \citep{APL2012b}.

\begin{figure}[h!]
\centering
\includegraphics[width=0.9\linewidth]{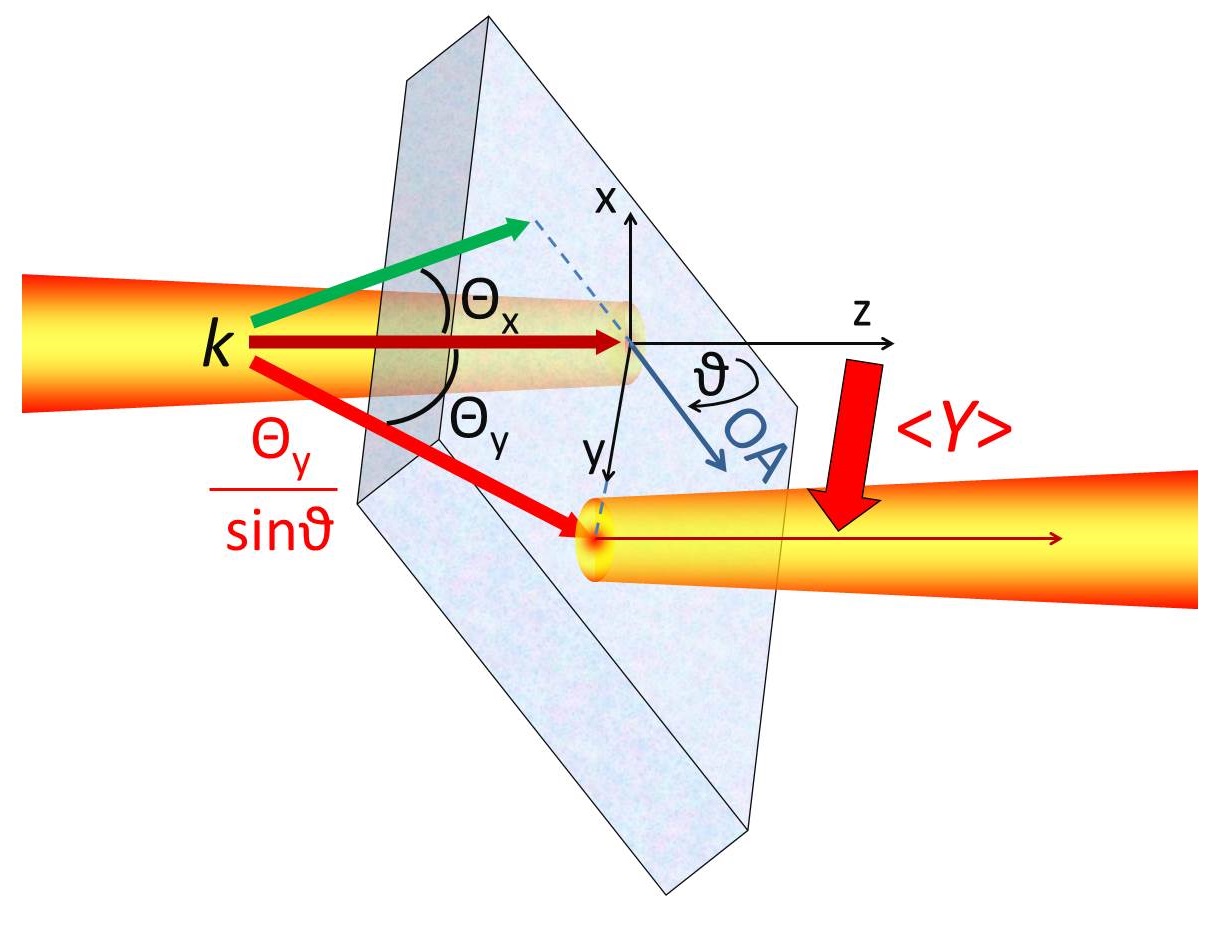}
\caption{3D geometry illustrating transmission of a paraxial beam through a tilted transparent polymer film. The beam experiences nano-meter scale transverse shift $\langle Y \rangle$  due to spin Hall effect in a polymer. The paraxial angles ($\Theta_{y},\Theta_{x}$) determine the direction of propagation of the wave vectors $k$ in the
incident beam. }
\label{fig:false-color}
\end{figure}

In this letter, we demonstrate experimentally that the spin-Hall effect of light and the transverse spin dependent
beam shift appears in the light transmission through a birefringent polymer with a tilted anisotropy axis as illustrated in Fig. 1. This new type of spin-Hall effect is
quite surprising for traditional optics, as has recently been demonstrated for a Quartz crystal plate \cite{Optica} because it implies weak
circular birefringence of a uniaxial crystal plate. To the best of our  knowledge, this is the first time this effect is reported in a polymeric material. Polymers are widely used as optical materials from transmission media to light sources with the advantages of low processing cost, mechanical flexibility, ease of large area fabrication and so on \cite{NaturePhoton2010}. Moreover, birefringent polymers are typically produced by electronic modulation, such as liquid crystals \cite{LC} or by stress induced mechanical effects \cite{OptExpress2017}. Therefore, this provides an external means of control for the birefringence induced in the polymer and could in turn be used as an optical switch in the nano-meter range, as opposed to natural crystals or metal where the birefringence if fixed by the parameters of the media.

The complete theory for light transverse shifts in uniaxial crystals was described in \cite{Optica}. For an input state described by a Jones vector $|\psi \rangle$ and a transmitted state described by Jones vector  $|\psi' \rangle$, the anisotropic transverse shift or spin-Hall effect of light is given by the expectation value of the position operator:

\begin{equation}
\langle Y \rangle= \langle \psi' | \hat{Y}| \psi' \rangle=\frac{cot(\vartheta)}{k}[-\sigma (1-\cos( \phi_0)) + \chi \sin(\phi_0) ],
\end{equation}

\noindent where $\phi_0$ represent the phase difference between ordinary and extraordinary wave propagating through the birefringent medium.

The spin-Hall effect can be measured either directly, via sub-wavelength shift [Eq. (1)] of the beam centroid \cite{23,24,25}, or via
various other methods including quantum weak measurements \cite{22,29,30,31,32,33,34}. The latter method allows for significant amplification
of the shift using almost crossed polarizers at the input and output
of the system, respectively. As before, the input polarizer corresponds to a pre-selected state $|\psi \rangle=(\alpha,\beta)^{T}$ (where $T$ stands for transpose operation), while the output polarizer corresponds to another post-selected polarization state $|\psi \rangle=(\alpha',\beta')^{T}$. The resulting beam shifts are determined by the weak value $\langle Y \rangle_{weak}$, instead of expectation values $\langle Y \rangle$, which can exhibit quantum weak amplification effect and lay outside the bounds of the spectrum of the operator. We analyze quantum weak amplification of the spin-Hall effect shift, considering an initial beam with $e$ polarization $|\psi\rangle=(1,0)^{T}$, while the post-selection polarizer is nearly orthogonal $|\phi \rangle=(\epsilon,1)^{T}, |\epsilon|<<1$. The weak value yields:

\begin{equation}
\langle Y \rangle_{weak}=\frac{1}{\epsilon k}\sin(\phi_0) \cot(\vartheta)+\frac{z}{z_{R}}\frac{1}{\epsilon k}(1-\cot(\phi_0))\cot(\vartheta),
\end{equation}

\noindent where $z_{R}$ is the Rayleigh length. The second (angular) term, becomes dominant in the far field zone, and presents weak amplification due to two reasons. Firstly, because $|\epsilon|<<1$, and secondly because $z >> z_{R}$. Note that the maximal achievable weak amplification at $|\epsilon| \approx (k \omega_0)^{-1}$ is of the order of the beam waist $\omega_0 z/z_{R}$.

To determine the classical beam shift we calculate the expectation value of the centroid displacement for a birefringent polymer, modeled as a tilted Quarz plate of thickness $d = 50 \mu m$. The phase difference can be expressed as:

\begin{equation}
\phi_0(\vartheta)=k[n_ {o} d_{o}(\vartheta)-\bar{n}_{e}(\vartheta)d_{e}(\vartheta)].
\end{equation}

Here, $n_{o}=1.54$  is the refractive index for the ordinary wave, $n_{e}(\vartheta)=n_{o}n_{e}/\sqrt{n_{e}\cos(\vartheta)+n_{o}\sin(\vartheta)}$ is the refractive index for
the extraordinary wave propagating at the angle $\vartheta$ to the optical
axis, $\bar{n}_{e}=n_{e}(\pi/2)=1.60$, and the distances of propagation of
the ordinary and extraordinary rays in the tilted plate are:

\begin{equation}
d_{e}(\vartheta)=\frac{\bar{n}_{e}(\vartheta)d}{\sqrt{\bar{n}_{e}^2(\vartheta) - \cos(\vartheta)^2}}, \hspace{0.3cm} d_{o}(\vartheta)=\frac{n_o d}{\sqrt{n_{o}^2-\cos(\vartheta)^2}}.
\end{equation}

Using Eqs. (1),  (2) and (3), in Fig. 2 (a) and (b)  we plot the phase difference 
and spin-Hall shifts (blue curves) as functions of
the tilt angle $\vartheta$. One can see that the transverse shift $\langle Y \rangle$   due
to the spin-Hall effect reaches wavelength-order magnitude, typical
for other spin-Hall systems in optics [20–24]. In contrast to
the IF shift in the reflection/refraction problems, here the transverse
shift as a function of $v$ displays two-scale behavior.
Namely, the fast oscillations in Fig. 2(b) originate from the term
$(1-\cos(\Phi_0)$), whereas the slow envelope corresponds to
the universal $\cot (\vartheta)$  factor in SOI terms.

\begin{figure}[h!]
\centering
\includegraphics[width=1\linewidth]{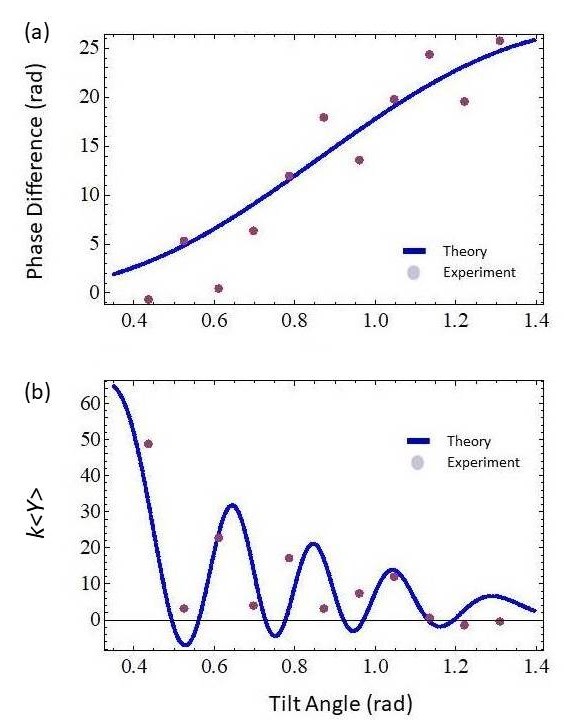}
\caption{Polarimetric measurement of (a) Phase difference $\Phi_{0}$ between ordinary and extraordinary polarizations,  (b)  adimensional transverse spin Hall  shift ($k \langle Y \rangle$), with wave-number $k=\frac{2 \pi}{\lambda}$. Theoretical prediction is indicated with blue curve and experimental data is indicated with purple dots. Agreement between experiment and theory is apparent.}
\label{fig:false-color}
\end{figure}

\begin{figure}[h!]
\centering
\includegraphics[width=\linewidth]{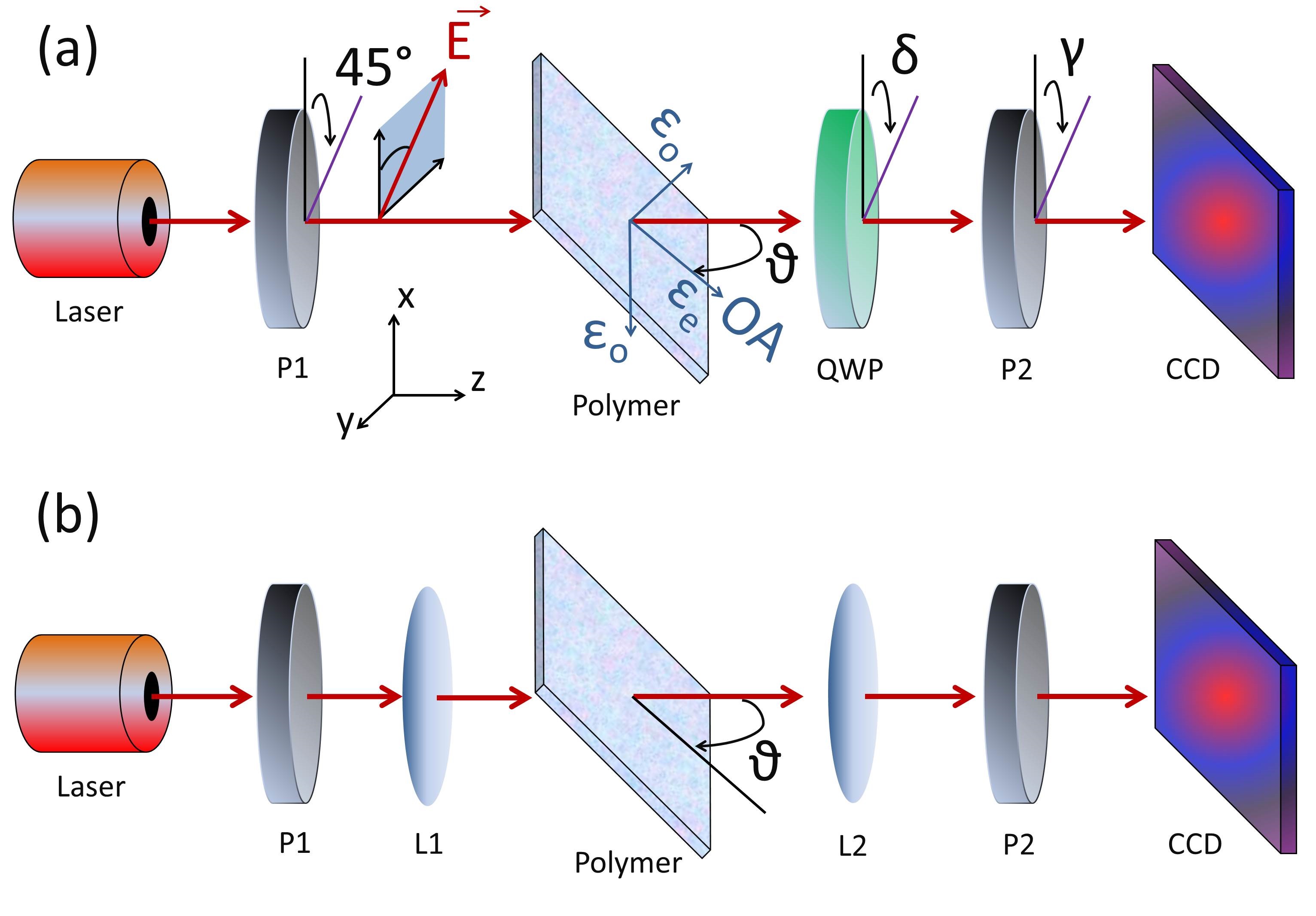}
\caption{
Schematic illustration of experimental setups for (a) polarimetric and (b) quantum weak measurements. P1 and P2 represent double Glan-Laser polarizers (Thorlabs GL10), QWP is quarter wave plate. Lenses are denoted as L1 and L2. Laser is He-Ne (Melles Griot Griot 05-LHR-111)  laser with the emission wavelength of 633 nm, the CCD camera model is Thorlabs WFS150-5C. The ordinary and extraordinary permittivities of the polymer film and their axes are denoted as $\varepsilon_{o}$ and $\varepsilon_{e}$, respectively.}
\label{fig:false-color}
\end{figure}


To verify the above theoretical predictions, we performed a series
of experimental measurements using the setups shown in Fig. 3.
We use a sample of free-standing birefringent polymer foil, similar to the type Newport 05RP32-1064. As a source of  incident Gaussian beam, we employed a He-Ne laser (Melles Griot Griot 05-LHR-111) of wavelength $\lambda=633$ nm. The laser radiation was
collimated using a microscope objective
lens. We measure the anisotropic phase difference $\Phi_0$  versus the angle of the tilt $v$ via Stokes polarimetry
methods \cite{28}. For this purpose we used the setup shown in Fig. 1(a).
The double Glan–Laser polarizer (Thorlabs GL10) (P1) selected the desired
linear-polarization state in the incident beam. In the first experiment,
this was 45° polarization, i.e., $\alpha=\beta=1/\sqrt{2}$. The beam then propagates through the polymer, and the Stokes parameters are measured using a Quarter Wave-Plate (QWP) at  a retardation angle $\delta$, and a second polarizer P2, with angle $\gamma$, as indicated in Fig. 3(a). The phase difference can be obtained via the Stokes parameters using the expression:

 \begin{equation}
 \phi_0=\tan^{-1} (\frac{S_3}{S_2}),
 \end{equation}

\noindent where $S_3=I(90^{\circ}, 45^{\circ})-I(90^ {\circ}, 135^{ \circ})$ is the normalized Stokes parameter for circular polarization, and $S_2=I(0 ^{\circ}, 45^ {\circ})-I(0^{ \circ}, 135^{\circ})$ is the normalized Stokes parameter in the diagonal basis, where the normalization factor $S_0$ is given by the total intensity of the beam. The measured phase using Eq. (5) is wrapped in the range $(-\pi, \pi)$. In order to determine the unwrapped phase difference we use an unwrapping algorithm \cite{Optica}, with a tolerance set to $0.001$ radians. The measured unwrapped phase difference is displayed in Fig. 2(a) (purple dots). The experimental spin Hall effect using Stokes polarimetry is shown in Fig. 2(b) (purple dots). The agreement between experiment and theory is apparent. We note that we observe a spin-Hall effect via Stokes polarimetry ($k<Y>$) using a $50 \mu m$ polymer film which is 10 times larger than the shift observed in Ref. \cite{Optica}, for a 1000 $\mu m$  Quartz sample. We ascribe this increase to the larger effective birefringence in the polymers \cite{OptExpress2017}.

Next, we performed the weak measurement of spin Hall shift, and observed quantum weak amplification effect using the quantum weak measurement setup in Fig. 3(b). The beam is imaged using a CCD camera (Thorlabs WFS150-5C). To this end we inserted two lenses (L1) and (L2) of focal distance $f=6$ cm. Polarizers Glan-Laser Polarizers P1 and P2 produce the pre-selected and post-selected states, with polarization states $|\phi\rangle$ and $|\psi\rangle$, respectively. The first lens (L1), of focal length 6 cm, produced a Gaussian beam with waist $\omega_0 = 30 \mu$m, and a  Rayleigh range $z_{R}=4.6$ mm. Therefore, for a CCD camera located at a distance $z=5$ cm the propagation amplification factor results $z/z_{R}=10.86$. The amplification factor due to crossed polarizers results $1/\epsilon \approx 1.83 \times 10^{-2}$. For $k=\frac{2 \pi}{\lambda}$, the overall weak amplification factor becomes $A = \frac{z}{z_{R}} \times  \frac{1}{k \epsilon}$ = 200, this is confirmed in the experiment where a displacement between centroids of $\Delta Y = 1000 \mu$m  between post-selection polarizers oriented  at $\epsilon=-1/1.83 \times 10^{-2}$ (Fig. 4(a)), and $\epsilon=+1/1.83 \times 10^{-2}$ (Fig. 4(c)) is measured at a tilt angle $\vartheta=20$ $^{\circ}$, thus amplifying the spin Hall effect by a factor $A = 200$. For crossed polarizers ($\epsilon=0$), the input Gaussian beam is split into a Hermite-Gaussian distribution (Fig. 4(b)), and the separation of the two peaks is approximately $\Delta Y=1000\mu m$ in $<Y>$.  

\begin{figure}[b!]
\centering
\includegraphics[width=0.8\linewidth]{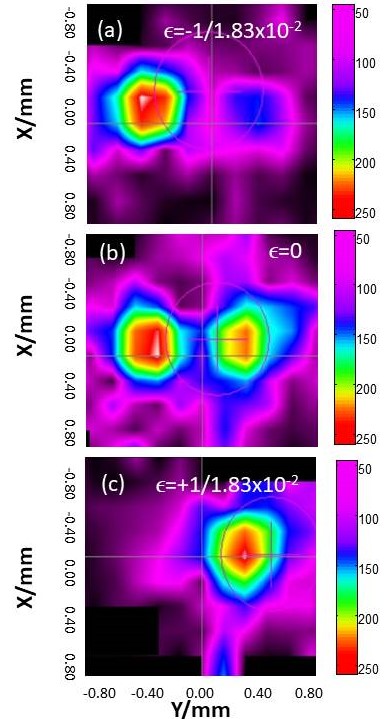}
\caption{Transverse intensity distributions (a.u.) for an  o-polarized beam transmitted through a tilted polymer plate and post-selected in the almost e-polarized state,  with a tilt angle $\vartheta=20^{\circ}$. (a) Post-selected polarization state with $\epsilon=-1/83 \times 10^{-2}$. The beam centroid is shifted, corresponding to a weak value measurement of $<Y>_{weak}=-500 \mu m$. (b) With crossed polarizers ($\epsilon=0$), the input Gaussian beam is split into a Hermite-Gaussian distribution with peaks separated by a distance $\Delta Y=1000\mu m$, which corresponds to a weak  amplification factor $A=200$. (c) Post-selected polarization state with $\epsilon=1/83 \times 10^{-2}$, corresponding to a weak value measurement of $<Y>_{weak}=+500 \mu m$ . Image in false color scale obtained with a CCD camera (Thorlabs WFS-150-SC).}
\label{fig:false-color}
\end{figure}

In conclusion, we demonstrated  experimentally the fine lateral circular birefringence of a tilted birefringent polymer, the first example of the spin-Hall effect of light in a polymer material. We reported experimental observations of this nano-meter scale effect using Stokes polarimetry techniques and quantum-weak-measurement techniques, reporting a quantum weak amplification factor of 200.  The birefringence in the polymer can be tuned using voltage in the case liquid crystals  or using mechanical stress in the case of stress-induced birefringence, therefore this lateral shift could be used as an optical switch in the nano-meter scale, opening the doors to a myriad of novel applications in photonics, nano-optics, quantum optics, and metamaterials. 

\vspace{-0.5cm}
\section*{}{\textbf{Funding.}}
Agencia Nacional de Promocion Cientifica y Tecnologica, PICT2015-0710 Startup, UBACyT PDE 2016 and UBACYT PDE 2017, Argentina; Villum Fonden DarkSILD project, Denmark (11116); Direkt{\o}r Ib Henriksens Fond, Denmark.
\vspace{-0.8cm}
\section*{}{\textbf{Acknowledgement.}} The authors are very grateful to Konstantin Bliokh and Ricardo Depine for insightful discussion.


\pagebreak


\end{document}